\documentclass[aps,prb,twocolumn,groupedaddress,showpacs]{revtex4}

\usepackage{color}
\usepackage{amssymb}
\usepackage{graphicx}
\usepackage{amsmath}
\usepackage{dcolumn}
\usepackage{bm}

\usepackage[colorlinks=true,linkcolor=blue,citecolor=blue,urlcolor=black]{hyperref}%
\expandafter\ifx\csname package@font\endcsname\relax\else
 \expandafter\expandafter
 \expandafter\usepackage
\expandafter\expandafter
 \expandafter{\csname package@font\endcsname}%
\fi

\begin{document}

\title{The Effects of vertical electric field and charged impurities on the spin-polarized transport of $\beta$-antimonene armchair nanoribbons}

\author{Shoeib Babaee Touski}
\affiliation{Department of Electrical Engineering, Hamedan University of Technology, Hamedan 65155, Iran}
\email{touski@hut.ac.ir}
\author{M. Pilar L\'opez-Sancho}
\affiliation{Instituto de Ciencia de Materiales de Madrid, CSIC, E-28049 Cantoblanco, Madrid, Spain}

\date{\today}

\begin{abstract}
The electronic properties of antimonene, single-layer Sb, are attracting great attention. 
In this paper, spin transport in armchair antimonene nanoribbon (ASbNR) is investigated. Following the tight-binding model we calculate both the transmission probability and the conductance by means of the non-equilibrium Green's function (NEGF) method. The effects of an external electric field vertical to the ribbon plane are explored. Our results indicate that the spin-flip rate increases with the vertical electric field. Disorder effects on spin transport are addressed by considering the presence of charged impurities. It is found that charged impurities also enhance the spin-flip rate but to a lesser extent than the out-of-plane electric field.
\end{abstract} 


\maketitle
\section{Introduction}
\label{intro}
Atomically thin two-dimensional materials have attracted great attention due to their excellent properties and promising applications in next-generation nanoelectronics devices. Furthermore, atomically thin two-dimensional (2D) materials offer an interesting playground for the study of novel electronic properties and have opened new routes to control their transport properties. Since the discovery of graphene, many new 2D materials have been synthesized, among them, elemental materials as arsenene and antimonene have been obtained confirming earlier theoretical predictions that mainly using density functional theory (DFT), had investigated Group-V derived 2D materials. Antimonene, a monolayer of Sb atoms \cite{zhang12topol,akturk15,wang2015atomically,rudenko17,wang2017many}, presents several phases with different structures. DFT results have shown that $\beta$-phase exhibits the best stability \cite{wang2015atomically} of all antimonene allotropes. In this phase, as in silicene the Si atoms, Sb atoms are arranged in a buckled honeycomb lattice (D$_{3d}$ point group) \cite{zhang2015atomically}.

The robust $\beta$-phase antimonene, theoretically predicted \cite{akturk2015single,wang2015atomically,pumera17} has been experimentally realized as a layer supported on various substrates \cite{yao2013evolution,kim2016,ji2016two,Takayama2014,Lei2016,Ares2016,FortinDeschnes2017} and by different methods \cite{ares2018recent} including liquid- and solid-phase exfoliation \cite{gibaja16,Ares2016}; epitaxial growth of a few layers on substrates has also been reported \cite{wu2017epitaxial,ji2016two,Lei2016,FortinDeschnes2017}. 
Based on both experimental and $ab$ $ initio$ modeling \cite{Ares2016} results, $\beta$-phase antimonene is considered a material with remarkable stability in air and water \cite{wu2017epitaxial,Ares2016}. This alone makes antimonene an appealing candidate for various applications because instability at ambient conditions is one of the main factors limiting the application of 2D materials \cite{kuriakose18,morishita15,kokabi2020electronic}. Antimonene exhibits many properties besides high stability, \cite{zhang2015atomically,Ares2016}, such as fast carrier mobility ($\mu_e=630$ and $\mu_h=1737 cm^2/V.s$) which exceeds that of monolayer MoS$_2$ and high electrical conductivity \cite{pizzi16,ji2016two} which make it a promising material for designing field-effect transistors (FETs)\cite{zhang15}.
Although Sb is a semimetal in the bulk the single layer antimonene is an indirect semiconductor with a broad range of bandgap from 0.76 eV \cite{wang2015atomically} to 2.28 eV \cite{singh16,zhang15} predicted theoretically.
The wide bandgap alongside the high mobility of the carriers favors this material for applications in metal-oxide-semiconductor field-effect transistors (MOSFETs)\cite{pizzi16}. In fact, quantum transport simulations give excellent performances of monolayer antimonene MOSFETs such as superior immunity against short channel effects and decent electron and hole mobilities \cite{pizzi16}.

A remarkable characteristic of these novel 2D materials is the tunability of their electronic properties. Calculations based on DFT have shown that tensile strain and a perpendicular electric field can tune the width as well as the indirect/direct characteristics of the bandgap \cite{shu2018electronic} and magnetic properties can be modulated by vacancies and strain\cite{Xing2019}.
Due to the Sb atomic weight, the intra-atomic spin-orbit interaction strength is significant and has to be considered for a correct description of the electronic structure and effective masses\cite{rudenko17} of the single-layers of antimony. Many potential applications of antimonene originate in this strong atomic spin-orbit coupling (SOC) as the spintronic response \cite{Lee2015,Soumyanarayanan2016} or the topological aspects and strain-driven band inversion\cite{Zhao2015}.

The sample geometry has important consequences on the properties of 2D materials, the ribbon geometry been among the most appealing for applications in nanoelectronics and novel devices. Antimonene nanoribbons have been synthesized by plasma-assisted process\cite{Tsai2016} opening new expectations about antimonene applications.
In general, nanoribbons with different edge and width show different electronic properties. In the case of antimonene armchair nanoribbons (ASbNRs) present an indirect bandgap while zigzag terminated nanoribbons have a direct bandgap as shown by DFT calculations of their electronic properties \cite{Wang2015elec}. Theoretical works on monolayer Sb nanoribbons of different terminations have shown the importance of external electric fields in the control of their electronic properties\cite{Song2017}; in nanoribbons with loss of inversion symmetry, the strong SOC of Sb induces significant spin splittings in valence band maximum and conduction band minimum and the applied electric field together with the strong SOC allows the control of the bandgap size and, in the zigzag terminated ribbons, eventually induces a band inversion thereby electric-field modulation of carrier compensation is achieved with direct consequences for the magnetoresistance effect\cite{Song2017}. 

 In this work, we study the effects of an out-of-plane electric field on the transport properties of ASbNRs.
 A theoretical work by Edo van Veen et al, [\onlinecite{van2018electronic}] studied the effect of an external electric field on the armchair and zigzag antimonene nanoribbon electronic structures, they found a decrease of the band gap and a valence band splitting due to the Rashba coupling induced by the lack of inversion symmetry accompanied by a different spin polarization of the two mini-valleys at both sides of the $\Gamma$ point at the Brillouin zone (BZ). We are especially interested in the spin-polarized electrical currents. We focus on ASbNRs and we investigate the effects of an external electric field on their spin-transport properties. We find that both the gap and the spin-polarized current of ASbNRs can be controlled by a vertical electric field.
 Impurities are known to affect the sample transport properties, especially in 2D materials, and some experimental works have reported that charged impurities are the main source of scattering in antimonene\cite{sarma2011electronic,bolotin2008ultrahigh}. We also investigate how charged impurities influence the spin-polarized transport in ASbNRs and we find that the potential amplitude and the density of impurities have a negative effect on the magnetoresistance and on the parallel conductance.

The remainder of the paper is structured as follows: in Section II we describe the model Hamiltonian, the out-of-plane electric field and impurity terms, and some details of the method of calculation. In Section III we present and discuss our results and finally, in Section IV, we present a summary and some conclusions. 
\section{Model and Method}\label{Sec:Approach}
\subsection{Thight Binding Hamiltonian}\label{Sec:TB}
Monolayer antimonene consists of Sb atoms in a buckled honeycomb structure with a two-atom unit cell. The in-plane lattice parameter is $a=1.12{\AA}$ and the two sublattices vertical displacement is $b=1.65{\AA}$. The valence and conduction bands are known to have a predominant $p$ character. We follow here the tight-binding model developed in Ref. \onlinecite{rudenko17} which includes six orbitals, three for each Sb atom of the unit cell. This Hamiltonian, obtained by a parametrization based on the formalism of maximally localized Wannier functions, gives an accurate description of the electronic structure of antimonene in a wide range of energy in agreement with first-principles results \cite{rudenko17}.The six Wannier functions per cell are a combination of the three $p$-orbitals centered on each Sb atoms, $|p_x \rangle $, $|p_y \rangle $ and $|p_z \rangle$, and can be expressed following reference \onlinecite{rudenko17} as:
\begin{equation}
\begin{bmatrix}
|p_1^K\rangle\\
|p_2^K\rangle\\
|p_3^K\rangle
\end{bmatrix}=M
\begin{bmatrix}
|p_x\rangle\\
|p_y\rangle\\
|p_z\rangle
\end{bmatrix}
\end{equation}
$M$ been a unitary matrix:
\begin{equation}
M=
\begin{bmatrix}
\frac{1}{2}(-1)^{K+1}\sin\alpha & \frac{\sqrt{3}}{2}\sin\alpha & (-1)^K \cos \alpha \\
\frac{1}{2}(-1)^{K+1}\sin\alpha & \frac{\sqrt{3}}{2}\sin\alpha & (-1)^K \cos \alpha \\
(-1)^{K}\sin\alpha & 0 & (-1)^K \cos \alpha
\label{Eq:orbitals}
\end{bmatrix}
\end{equation}
where K is the sub-lattice index (1 or 2), and $\alpha\approx55.3^{\circ}$ is the angle between an orbital from the z-direction. 

The effective Hamiltonian based on these three orbitals can be written as:
\begin{equation}
\label{Eq:hamiltonian}
\hat{H_0}= \sum_{\left\langle i,j \right\rangle;l,m} t_{i,j;l,m}\hat{c}_{i;l}^\dagger \hat{c}_{j;m}+\hat{H}_\mathrm{SO}
\end{equation}
where $i,j$ are the atomic position indices, $l$ and $m$ label the atomic orbitals, $\hat{c}_{i;l}^\dagger (\hat{c}_{i;l})$ creates (annihilates) an electron at orbtial $l$ of site $i$ and $t_{i,j;l,m}$ are hopping parameters. We consider the $t_{i,j;l,m}$ amplitudes given in Ref \onlinecite{rudenko17}. $\hat{H}_\mathrm{SO}$, the SOC Hamiltonian, in terms of the $p$-orbital basis, can be expressed as\cite{Chaghazardi2016}:
\begin{equation}\label{Eq:HSO}
\hat{H}_\mathrm{SO}= \sum_{i;l,m}\frac{\lambda_{i;l}}{\hbar} \left[ \frac{\hat{L}_{i;l}^+ \hat{S}_{i;m}^- + \hat{L}_{i;l}^- \hat{S}_{i;m}^+}{2} + \hat{L}_{i;l}^z \hat{S}_{i;m}^z \right]
\end{equation}
where $\lambda$ is the intra-atomic SOC constant (we consider $\lambda=0.34eV$ consistent with the intraatomic stregth of neutral Sb atoms\cite{rudenko17}) and ${\hat{\cal O}}^{\pm}={\hat{\cal O}}^{x}\pm i{\hat{\cal O}}^{y}$ are the standard ladder operators, with ${\hat{\cal O}}={\hat L},{\hat S}$. We can distinguish two different contributions to the SOC Hamiltonian (\ref{Eq:HSO}), the first term which leads to spin-flip processes, and the spin-conserving term $\propto \lambda \hat{L}^z\hat{S}^z$. 
 The SOC Hamiltonian can be transformed to the Wannier bases set following the equation:
\begin{equation}
H_{SO,w}=MH_{SO,p}M^{-1}
\end{equation} 
Where $H_{SO,p}$ represents the Hamiltonian matrix elements for $p$-orbitals and $H_{SO,w}$ for Wannier functions.

\subsection{Vertical electric field and charged impurites}\label{Sec:EF}

To investigate the effects of out-of-plane voltages on the monolayer antimonene we add an on-site term to the Hamiltonian. The electric field can be modeled by adding the term:
\begin{equation}
H_{E}=e\varepsilon_{ext} d,
\end{equation}
where $e$ is the unit charge, $\varepsilon_{ext}$ is the vertical electric field and $d$ is the buckling height of antimonene. 
Due to the buckling, the vertical electric field will result in different potentials on the two different sub-lattices.

Charged impurities are modeled adding a potential amplitude to a number $N_{imp}$ of lattice sites chosen at random from the total $N$ sites, therefore the density of scatterers is $n_{imp} = N_{imp} /N$. The potential amplitude $U_n$ at the $n$ site belongs to a distribution of random values within the interval $[-\delta U,\delta U]$. Since the impurity potential would be screened in a sample, we will smooth the $U_n$ over the correlation length $\xi$, considering\cite{rycerz07,lewenkopf08,touski2020comparative}:  
\begin{equation}\label{Eq:U}
U_{imp}(r)=\sum_{n=1}^{N_{imp}} U_n exp\left(-\frac{|r-R_n|^2}{2\xi^2}\right),
\end{equation}
 To model charged-impurity disorder, a superposition of $N_{imp}$ Gaussian potential fluctuations is added to the Hamiltonian as diagonal terms.

\begin{figure}[t]
	\centering
	\includegraphics[width=1.0\linewidth]{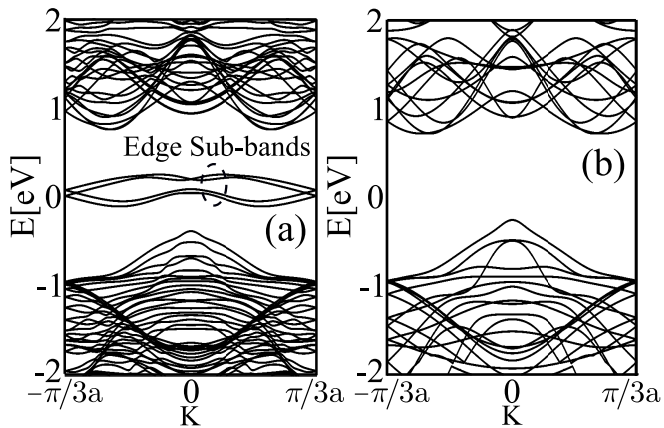}
	\caption{Band structure for armchair antimonene nanoribbon of width nW=18 for (a) non-periodic and (b) periodic boundary conditions. }
	\label{fig:fig1}
\end{figure}

\subsection{Non-equilibrium Green's function method}\label{Sec:NE}
The non-equilibrium Green's function (NEGF) formalism \cite{pourfath14non} is used to study spin transport in armchair antimonene nanoribbons. The channel and the two contacts are assumed to be of the same material. A detailed description of the NEGF with spin consideration has been previously given \cite{touski2017}.

 The transmission probability is calculated from
\begin{equation}
T(E)=trace\left[\underline{\Gamma}^L \underline{G^r} \underline{\Gamma}^R \underline{G^a}\right],
\end{equation}
where $\underline{\Gamma}^{L,R} = i\left(\underline{\Sigma}^{L,R} -\left( \underline{\Sigma}^{L,R}{}\right)^\dagger\right)$
are the broadening of the two contacts. The total transmission is given by:
\begin{equation} 
\begin{split}
T_{\sigma\sigma'}(E) &= trace \left[ \Gamma^L_{\sigma\sigma}G^r_{\sigma\sigma'}\Gamma^R_{\sigma'\sigma'}G^a_{\sigma'\sigma} \right],
~~~~\sigma,~\sigma'=\uparrow,\downarrow.
\end{split}
\label{Eq:fourTE}
\end{equation}

 Four spin-dependent transmission probabilities are calculated from up and down spin at the left contact to up and down spins at the right one.
$T_{\uparrow\uparrow}(E)$ and $T_{\downarrow\downarrow}(E)$ represent parallel spin transmission, and $T_{\uparrow\downarrow}(E)$ and $T_{\downarrow\uparrow}(E)$ antiparallel, spin-flip, transmission.


\begin{figure}
	\centering
	\includegraphics[width=1\linewidth]{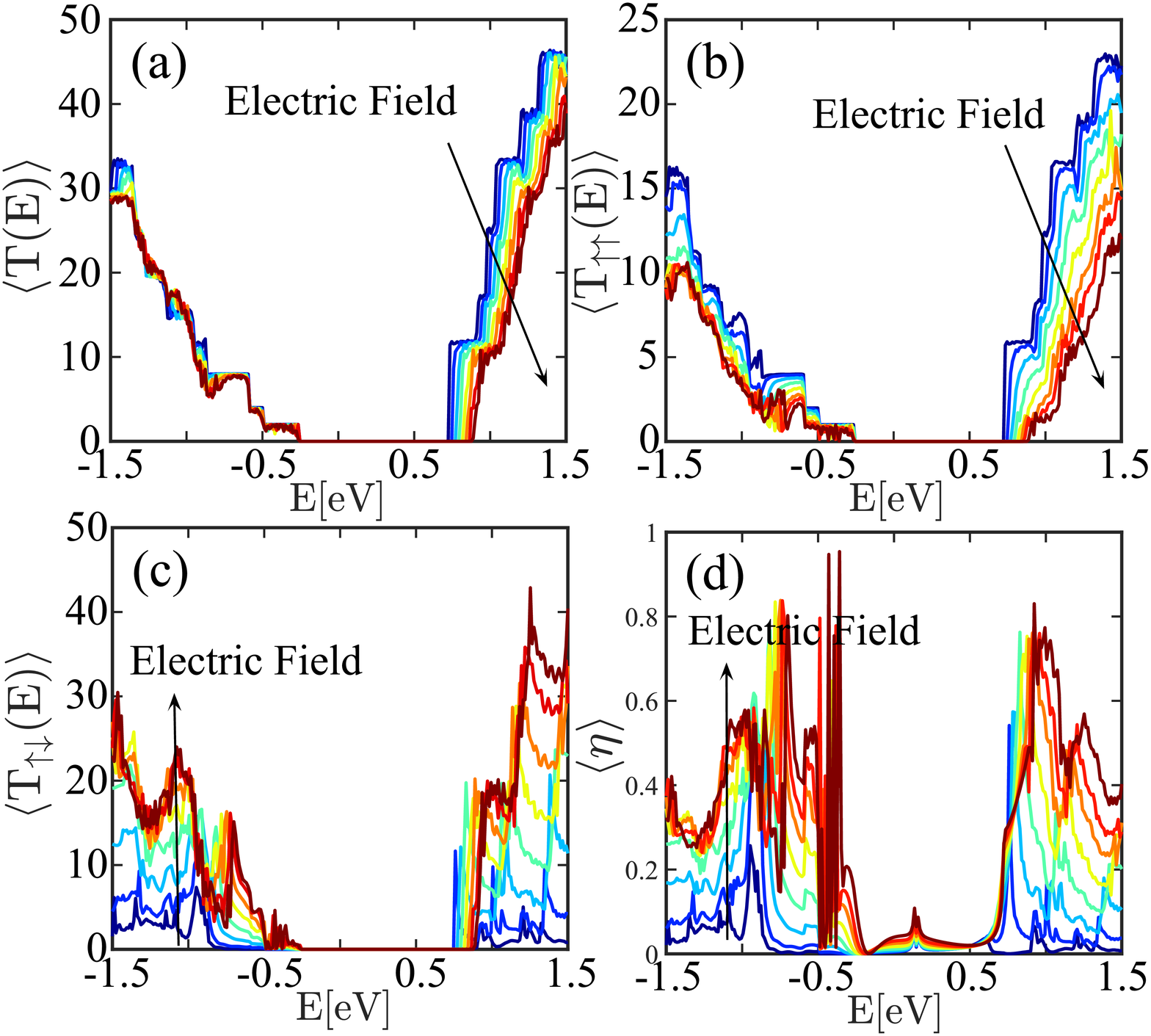}
	\caption{(a) Total transmission, (b) parallel spin transmission $T_{\uparrow\uparrow}$, (c) antiparallel transmission $T_{\uparrow\downarrow}$ and (d) spin-flip efficiency $\eta$ as a function of energy for different values of the electric field, for a ribbon of $\mathrm{L=10nm, nW=18}$.}
	\label{fig:fig2}
\end{figure}

\section{Results and discussion}
\label{Result}

\begin{figure*}
	\centering
	\includegraphics[width=1.0\linewidth]{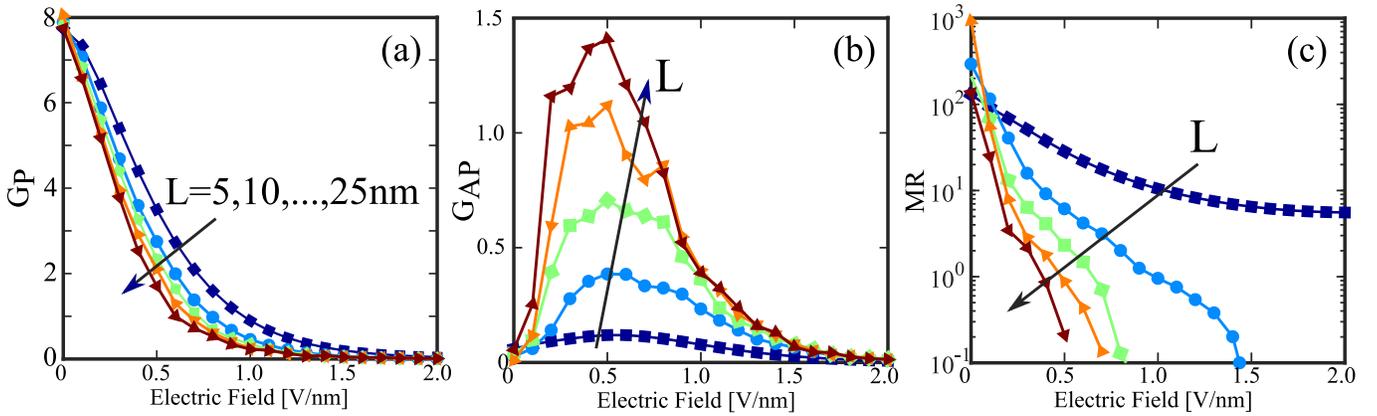}
	\caption{ Parallel (a) and anti-parallel (b) conductances, and magneto-resistance (MR) as a function of external electric field for different channel lengths.}
	\label{fig:fig3}
\end{figure*}

The band structure of an antimonene armchair nanoribbon (ASbNR) of width $nW=18$ (number of atoms in the transverse direction) obtained from the TB+SOC Hamiltonian [\ref{Eq:hamiltonian}] is shown in Fig.\ref{fig:fig1}(a). The mid-gap states correspond to edge states mainly localized at the edge atoms. The edge states disappear when periodic boundary conditions are considered in the direction perpendicular to the armchair edge, as seen in Fig.\ref{fig:fig1}(b). These band structures, obtained from the Hamiltonian expressed in Eq.(\ref{Eq:hamiltonian}) present good agreement with those obtained from first-principles calculations as confirmed in Refs.\onlinecite{rudenko17} and \onlinecite{van2018electronic}. The periodic ASbNR band structure shown in Fig.\ref{fig:fig1}(b), similar to that of a monolayer antimonene, corresponds to an indirect semiconductor with band gap $E_g=0.985eV$ close to the gap of pristine monolayer antimonene. The valence band maximum (VBM) lies at $\Gamma$ while the conduction band minimum (CBM) is along the $\Gamma-M$ line. As in a single-layer antimonene, valence and conduction bands are not symmetric and the effective masses of electron and holes are different.
Our calculations are carried out considering periodic boundary conditions in the transverse direction, therefore the midgap states are not present. These states are highly localized at the edges. At zero bias, these states are spin degenerate, the application of an external electric field breaks the spin degeneracy and the two lower midgap bands move down while the two higher bands move up \cite{van2018electronic}.

\subsection{Electric field}

As stated above Sb atoms have a strong SOC thus suggesting that antimonene is a good candidate for field-effect transistors and offers the possibility of controlling spins by applying electric fields. We will investigate the effect of an external electric field perpendicular to the ASbNR on its electronic properties. The application of a perpendicular bias voltage induces a potential difference between the two sublattices of Sb atoms, due to the buckled honeycomb lattice structure of antimonene. The spin-dependent transmissions calculated are plotted in Fig. \ref{fig:fig2} as a function of energy for different values of the external electric field. The total transmission is shown in Fig. \ref{fig:fig2}(a) is the sum of the four spin-dependent transmissions and, although the conduction band transmission shifts to higher energies when the electric field increases and the bandgap increases, the total transmission does not change significantly with the electric field.

 The spin-conserving transmission $T_{\uparrow\uparrow}$ decreases with increasing external electric field, as can be observed in Fig. \ref{fig:fig2}(b). On the other hand, the spin-flip transmission $T_{\uparrow\downarrow}$ increases with the electric field, as shown in Fig. \ref{fig:fig2}(c) where the evolution of the $T_{\uparrow\downarrow}$ with the electric field is plotted. While the maximum of the valence band is shifted to lower energies, the minimum of the conduction band moves up to higher energies, thus the bandgap widens with the perpendicular electric field. The spin-orbit coupling is enhanced by the electric field, increasing the spin-flip processes and increase of the transmission from up(down) spin to down(up) spin. Therefore, $T_{\uparrow\downarrow}$ and $T_{\uparrow\uparrow}$ present opposite evolution with the electric field. In Fig. \ref{fig:fig2}(d) we represent the evolution with the field of the spin-flip relative transmission or spin-flip efficiency defined as \cite{michetti10}:
\begin{equation}
\eta=\frac{T_{\uparrow\downarrow}}{T_{\uparrow\uparrow}+T_{\uparrow\downarrow}}
\end{equation}
which gives a measure of the efficiency of spin control. Spin-flip efficiency $\eta$ increases with the electric field as can be seen in Fig. \ref{fig:fig2}(d).


Next we compute the linear-response conductance within the Landauer theory:
\begin{equation}
G_{\sigma\sigma'}=\frac{e^2}{h}\int T_{\sigma\sigma'}(E)\left(-\frac{\partial f(E)}{\partial E}\right) dE
\end{equation}
where parallel and anti-parallel conductances are given by: 
\begin{eqnarray}
\label{Eq:Ga-Gap}
\nonumber
G_P = G_{\uparrow\uparrow}+G_{\downarrow\downarrow} \\
G_{AP} = G_{\uparrow\downarrow}+G_{\downarrow\uparrow}.
\end{eqnarray}
The magnetoresistance (MR) is an important magnitude for spintronic devices and can be calculated from the conductance:
\begin{equation}
MR=\frac{R_{AP}-R_P}{R_P}=\frac{G_P-G_{AP}}{G_{AP}}
\end{equation}
Once the transmission probability is obtained by solving the NEGF equation, parallel and anti-parallel conductances are calculated using Eq. (\ref{Eq:Ga-Gap}). Finally, the MR is calculated from these conductances.

The parallel and anti-parallel conductances, calculated in the linear regime, are plotted in Fig. \ref{fig:fig3}. Parallel conductance decreases with increasing electric field. Since the bandgap increases with the electric field, the concentration of electrons decreases, and the conductance also decreases as can be observed in Fig. \ref{fig:fig3}(a) for channels of different lengths from 5 to 25nm. However, the behavior of the anti-parallel conductance, shown in Fig. \ref{fig:fig3}(b) is different, first it increases with the electric field then, after reaching a maximum value, it decreases even if the electric field continues to increase. This behavior is due to the increase of spin-flip processes for the low electric field that enhances the anti-parallel conductance. From a given value of the electric field, the electron concentration decreases with the field as the bandgap widens. The decrease in electron concentration dominates the spin-flip conductance that also decreases.

The electric field effects on the magneto-resistance can be seen in Fig. \ref{fig:fig3}(c) where the MR is plotted, for different channel lengths, as a function of the electric field. For an ASbNR of $L=10nm$, the large MR of 1000 decreases rapidly to $10^{-1}$ when applying an electric field of $\mathrm{0.7V/nm}$ This reduction of the MR with the external electric field suggests that spin-flip processes can be efficiently controlled in this 2D material by applying an electric field.
The influence of the channel length can also be observed in Fig. \ref{fig:fig3}. While the parallel conductance experiences only small changes with the length of the channel (see panel (a)) the antiparallel conductance undergoes a significant increase (see panel (b)) due to the increase of the spin-flip processes with the channel length. The increase in the antiparallel conductance $G_{AP}$ induces a decrease in the $MR$. Therefore, the channel length significantly affects the behavior of the $MR$ as shown in panel (c). The decrease of the MR is greater the longer the channel as can be seen in the slope of the curves: an electric field of $2 V/nm$ reduces the MR of 100 twenty times for a channel of $L=5nm$ while, for a channel of $L= 25nm$ the MR decreases four orders of magnitude to $10^{-1}$ by applying a field of only $0.5 V/nm$

\begin{figure}[t!]
	\centering
	\includegraphics[width=1\linewidth]{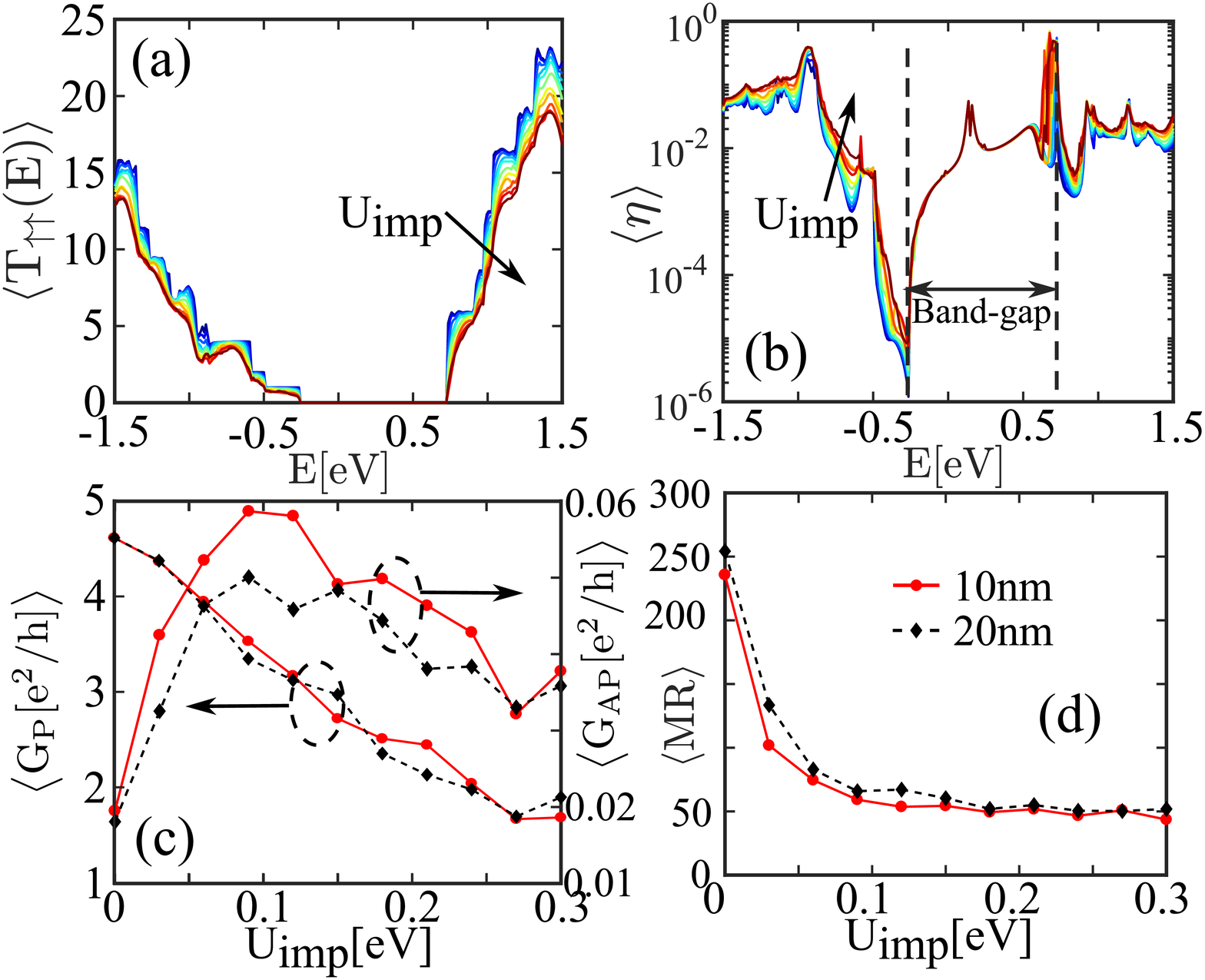}
	\caption{ $T_{\uparrow\uparrow}$ (a) and $\eta$ (b) as a function of energy for different charged impurities amplitudes. Parallel and anti-parallel conductances (c) and MR as a function of charged impurity amplitudes (d) for $L=10nm$ and $L=20nm$ channel lengths. Results for $nW=18$. $\mathrm{n_{imp}=0.1}$ and $\mathrm{\xi=3a}$.}
	\label{fig:fig4}
\end{figure}

\subsection{Charged impurities}
Real samples always present disorder of different types that affects their electronic properties and charged impurities constitute one of the most common types of disorder. In order to investigate the effects of charged impurities, we generate ensembles of disorder realizations for each of the randomly chosen potential amplitudes. Considering $U_{imp}$ which models the screened charge amplitudes according to Eq.(\ref {Eq:U}) with a correlation length $\xi$, at random $N_{imp}$ sites of the ribbon, we calculate the average transmissions. The results are represented in Fig. \ref{fig:fig4}. The average $T_{\uparrow\uparrow}$ slightly decreases with increasing impurity potential amplitude as can be seen in Fig. \ref{fig:fig4} (a) where $\delta U$ varies from 0 to $0.3 eV$. This behavior has also been obtained in Silicene nanorribons\cite{touski19spin}. On the contrary, the average spin-flip efficiency increases with $U_{imp}$ as shown in panel (b) because the presence of charged impurities increases spin-flip processes. Due to the scattering of electrons by charged impurities, the parallel or spin-conserving conductance $G_{P}$ is reduced, while the antiparallel conductance $G_{AP}$ increases for low potential amplitudes of the charged impurities, reaches a maximum, and then decreases for higher potential amplitudes. The behavior of $G_{P}$ and $G_{AP}$, as shown in panel (c), is similar for the two different channel lengths, $L=10nm$ and $L=20nm$. The magnetoresistance is represented in Fig. \ref{fig:fig4}(d) versus the potential amplitude, MR decreases for potentials lower than $\mathrm{\delta U=0.1eV}$ and then stabilizes at 50 for increasing values of $\delta U$, showing the same behavior for the two channel lengths.

\begin{figure}
	\centering
	\includegraphics[width=1\linewidth]{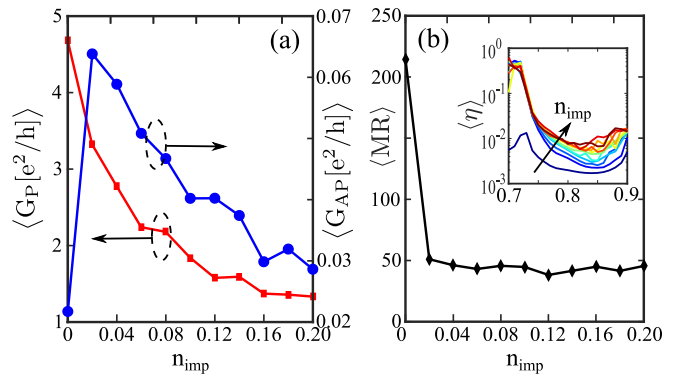}
	\caption{ Parallel and anti-parallel conductances as a function of $n_{imp}$ (a). MR versus impurity density (b); the inset shows the spin-flip efficiency of the conduction band for different values of the impurity density $n_{imp}$. Results for $ nW=18$, $L=10nm$, $\mathrm{\delta U=0.3eV}$ and $\mathrm{\xi=3a}$. }
	\label{fig:fig5}
\end{figure}

We have studied the influence of the density of impurities $n_{imp}$, i.e., the relation of impurities with respect to the total number of atoms of the ribbon, in the parallel and antiparallel conductances, in the range of $n_{imp}$ from 0 to 0.2. In Fig.\ref{fig:fig5} (a) it is shown that the average $G_{P}$ decreases as the density of impurities increases as $G_P \propto 1/n_{imp}$. The $G_{AP}$, after a sharp increase for low densities, decreases as the density of charged impurities increases. The behaviors shown by $G_{P}$ and $G_{AP}$ with respect to the density of impurities are qualitatively similar to what they show versus the impurity potential amplitudes. The magnetoresistance plotted in Fig.\ref{fig:fig5} (b) as a function of the density of impurities, experiences a sharp reduction for low densities of impurities and remains without significant variations from $n_{imp}\approx 0.4 cm^{-2}$. The evolution of the spin-flip efficiency with the impurity density is shown in the inset of Fig. \ref{fig:fig5}, it can be observed that it increases with the density of impurities.

\section{conclusion}
\label{Sec:Conclusions}
We have investigated the effects of a perpendicular electric field and of charged impurities on the transport properties on armchair Sb nanoribbons, by means of a tight-binding model including spin-orbit interactions. We have considered periodic boundary conditions in both directions, in order to avoid the presence of edge states. We have found that the electric field enhances the bandgap and reduces the total and spin parallel transmissions. The antiparallel transmission and the spin-flip efficiency increase with the external electric field. 

Our results show that the length of the ribbon significantly affects the conductance. The parallel conductance decreases slightly with the length of the channel, but the antiparallel conductance experiences a strong increase due to the raise of spin-flip processes with the length of the channel, therefore the MR decreases.
In the presence of charged impurities, the parallel transmission is reduced when the potential amplitude increases while the spin-flip efficiency grows up. Both the parallel conductance and the magnetoresistance decrease when the potential amplitude of the impurities increases. The antiparallel conductance increases for low amplitudes and then decreases. The evolution of the $G_{P}$, $G_{AP}$ and $MR$ with the density of impurities is similar to that presented with the potential amplitude. In this case, the effects of the channel length on the conductance and the magnetoresistance are weak. 

The characteristics of ASbNRs, buckled geometry, and strong SOC, cooperate so that an electric field is sufficient to create and control spin-polarized currents. This is an interesting result which opens new paths for the inclusion of ASbNRs in spintronics devices. 

\section{aknowledgements}
The authors gratefully acknowledge helpful discussions with R. Rold\'an.
This research has been supported in part by the Spanish MINECO Grant No. PGC2018-099199- B-I00, the European Union Structural Funds, and the Comunidad de Madrid NMAT2D-CM Program (Grant No. S2018-NMT-4511).


\end{document}